\documentclass{pasj00}
\draft

\begin{document}
\SetRunningHead{Deguchi et al.}{SiO Masers Survey in the Galactic Center}
\Received{2003/10/22}
\Accepted{2003/}

\title{SiO Maser Survey of the Large-Amplitude Variables in the 
Galactic Center}

\author{Shuji \textsc{Deguchi}$^{1}$, Hiroshi \textsc{Imai}$^{2, 3, 4}$, 
        Takahiro \textsc{Fujii}$^{2,4}$, Ian S. \textsc{Glass}$^{5}$, 
Yoshifusa \textsc{Ita}$^{6}$,}
\author{Hideyuki \textsc{Izumiura}$^{7}$, Osamu \textsc{Kameya}$^{2, 8}$, 
        Atsushi \textsc{Miyazaki}$^{1}$, Yoshikazu \textsc{Nakada}$^{6}$
}

\and
\author{Jun-ichi {\sc Nakashima}$^{9, 10}$}

\affil{$^{1}$ Nobeyama Radio Observatory, National Astronomical Observatory,\\
              Minamimaki, Minamisaku, Nagano 384-1305}
\affil{$^{2}$ VERA Project Office, National Astronomical Observatory, 
              2-21-1 Osawa, Mitaka, Tokyo 181-8588}
\affil{$^{3}$ Joint Institute for VLBI in Europe, 
              Postbus 2,  7990 AA Dwingeloo, The Netherlands}
\affil{$^{4}$ Faculty of Science, Kagoshima University, 
              1-21-35 Korimoto, Kagoshima 890-0065}
\affil{$^{5}$ South African Astronomical Observatory,
              PO Box 9, Observatory 7935, South Africa}
\affil{$^{6}$ Institute of Astronomy, School of Science, The University of Tokyo,
              2-21-1 Osawa, Mitaka, Tokyo 181-0015}
\affil{$^{7}$ Okayama Astrophysical Observatory, National Astronomical 
Observatory, \\ Kamogata, Asakuchi, Okayama 719-0232}
\affil{$^{8}$ Mizusawa Astrogeodynamics Observatory, National 
Astronomical Observatory, Mizusawa, Iwate 023-0861}
\affil{$^{9}$ Department of Astronomical Science, The Graduate University 
for Advanced Studies,\\
Minamimaki, Minamisaku, Nagano 384-1305}   
\affil{$^{10}$ Department of Astronomy, University of Illinois at Urbana-Champaign\\
1002 W. Green St. , Urbana, IL 61801-3074, U.S.A.}

 \author{\\(PASJ  April 28 issue in press--- Version: 2004/02/27)\\}


%


\maketitle

\begin{abstract}

We have surveyed $\sim 400 $ known large-amplitude variables within 15$'$ of
the galactic center in the SiO $J=1$--0 $v=$ 1 and 2 maser lines at 43 GHz,
resulting in 180 detections.  SiO lines were also detected from 16 other
sources, which are located within 20$''$ (the telescope half beamwidth) of the
program objects.  The detection rate of 48 percent is comparable to that
obtained in Bulge IRAS source surveys. Among the SiO detections, five stars
have radial velocities greater than 200 km s$^{-1}$. The SiO detection rate
increases steeply with the period of light variation, particularly for stars
with $P>500$ d, where it exceeds 80\%.  We found that, at a given period,
the SiO detection rate is approximately three times that for OH.  These facts
suggest that the large-amplitude variables in the Nuclear Disk region are
AGB stars similar in their overall properties to the inner and outer Bulge
IRAS/SiO sources. From the set of radial velocity data, the mass
distribution within 30 pc of the galactic center is derived by a new method
which is based on the collisionless Boltzmann equation integrated along the
line of sight. The mass within 30 pc is about $6.4 [\pm 0.7] \times 10^7 $
M$_{\odot}$ and the mass of the central black hole is $2.7 [\pm 1.3] \times
10^6 $ M$_{\odot}$. Consideration of the line-of-sight velocity of each star
and its potential energy leads to the conclusion that the five high-velocity
stars come from galactocentric distances as high as 300 pc. The
high-velocity subsample of stars with negative radial velocities exhibits a
tendency to have brighter $K$ magnitudes than the subsample of stars
with positive velocities. The origin of these high-velocity stars is
discussed.

\end{abstract}

\section{Introduction}

Radial-velocity data concerning stellar maser sources are useful for
studying the dynamical behavior of the central part of the Galaxy
(\cite{lin92a}; \cite{izu95}; \cite{sjo98}; \cite{deg00}). At visible
wavelengths, they are difficult to obtain in the galactic center region
because of interstellar extinction. Instead, most information comes from
radio or near-infrared observations [for example, \citet{sel87}]. In
particular, observations of SiO and OH masers give radial velocities of
stars accurate to within a few km s$^{-1}$. The masers arise in the
circumstellar envelopes of mass-losing stars on the Asymptotic Giant Branch
(AGB), which are intrinsically bright in the near- and mid-infrared regions,
and which can potentially be identified at these wavelengths. 
Large numbers of candidate stars suitable for pointed maser surveys 
toward the nuclear disk have
been discovered in the near-infrared $K$ band by making use of their
characteristic large-amplitude variability (\cite{gla01}). 

The dynamical behavior of the central region of the Galaxy has attracted
much attention, especially in relation to the central black hole (for
example, \cite{mor96}).  Proper motions of stars have been measured in the
near-infrared $K$ band near the black hole (\cite{gen00}; \cite{ghe00}), and
proper motions of SiO maser stars have also been measured within the central
15$''$ (\cite{rei03}). These were used to find the position of Sgr A* in the
$K$-band images (\cite{men97}).  This paper concentrates on the dynamics of
stars located towards the outer part of the central star cluster around the
black hole, i.e., at about 2--30 pc distance, where the gravitational force
of the black hole ceases to influence the stellar motions, and the
stellar system is nearly self-gravitating. The dynamical (rotational) time scale in this
region is a few $\times 10^6$ y, while the ages of the AGB stars are
$10^7$ -- $10^9$ y. Therefore, these stars are considered to be dynamically
well relaxed (e.g., \cite{hoz00}). Since the bar-like structure of the Galactic bulge was
discovered (\cite{bli91}; \cite{nak91}; \cite{dwe95}), it has been
recognized that non-circular motions must be taken into account when
interpreting observational data such as the CO gas distribution in the
central nuclear disk (\cite{bin91}; \cite{wei99}).  Because double bars and nuclear rings
have been proposed as efficient mechanisms for feeding gas into the centers
of galaxies (\cite{shl89}), it has become additionally important to look for
signs of non-circularity in the motions of gas and stars.

In this paper, we report on the results of an SiO maser survey of Large
Amplitude Variables (Miras or semiregulars; abbreviated as LAV hereafter) in
a $24' \times 24'$ area of the galactic center (\cite{gla01}), whose
amplitudes and periods are known (\cite{woo98}; \cite{gla01}). 
Because these
stars are located at approximately the same distance (about 8 kpc) from the
Sun, they constitute an ideal sample for studying the statistical
characteristics of AGB stars and their detectability in the maser lines. In
addition, surveying these sources gives accurate radial velocities, and
provides basic data for investigating the kinematics of the galactic nuclear
disk.  Although the mass within this region has been obtained previously
by various methods [for example, \citet{lin92a}], none of them 
[except \citet{sah96}] are fully valid for treating the problem; 
the results obtained are likely to be in
error by a factor of a few between the radii of 2 and 30pc. In this paper,
we analyze the new SiO radial-velocity data set using the Boltzmann equation
integrated along the line of sight. We also consider the origin of the
high-velocity stars seen toward the galactic center.



\section{Observations}

Simultaneous observations in the SiO $J=1$--0, $v=1$ and 2 transitions at
42.122 and 42.821 GHz, respectively, were made with the 45-m radio telescope
at Nobeyama during the periods of 2001 February--May, 2002 March--May, and
2003 May.  We used a cooled SIS mixer receiver (S40) for the 43 GHz
observations and accousto-optical spectrometer arrays, AOS-H and AOS-W,
having bandwidths of 40 and 250 MHz. The effective velocity resolution of
the AOS-H spectrometer is 0.3 km s$^{-1}$. They cover the velocity range of
$\pm 350 $ km s$^{-1}$, for both the SiO $J=1$--0 $v=1$ and 2 transitions,
simultaneously. The overall system temperature was between 200 and 300 K,
depending on the weather condition. The half-power telescope beam width
(HPBW) is about 40$''$. The antenna temperature ($T_{a}$) 
given in the present paper is that corrected for the atmospheric 
and telescope ohmic loss but not for the beam or aperture efficiency.
The conversion factor of the antenna temperature to the flux density is 
about 2.9 Jy K$^{-1}$. To save observation time, we employed the position
switching sequence, Off--On1--On2--On3, where the off position was taken 7
arc minutes west of the first-object position (On1) in right ascension; the
separation of the off position corresponds to the angle moved by an object
in the sky during the typical integration (20 s), additional
telescope-slewing, and settling time (10 s), so that the integrations were
made at nearly the same elevation angle.  With this sequence, we observed
three stars at once and saved about 50\% of the total observation time. Further
details of SiO maser observations using the NRO 45-m telescope have been
described elsewhere (\cite{deg00}), and are not repeated here.

The sources observed in this paper are the large-amplitude variables (LAVs)
in a $24'\times24'$ area of the galactic center (\cite{gla01}).  The
positions of the observed sources are shown in Figure 1. In this sample, the
star name was designated by field and object numbers, e.g., 3--49.  The
total number of the objects listed in \citet{gla01} was 418. Note that,
however, a star near the edge of a field was sometimes given a second name
because it appeared in an adjacent field. We counted 15 such objects; the
measured positions of these stars are coincident with a few arcsec accuracy.
In field 11, the positions turned out to be shifted by about 30$''$ and
corrected values were listed in an erratum (\cite{gla01}); note that the
latter includes an additional variable, 11-307. Additionally, three LAV
stars, which were identified later with the help of ISOGAL observations
(\cite{omo03}), were observed. The positions of these 3 stars are given in
table 1.


\begin{figure}
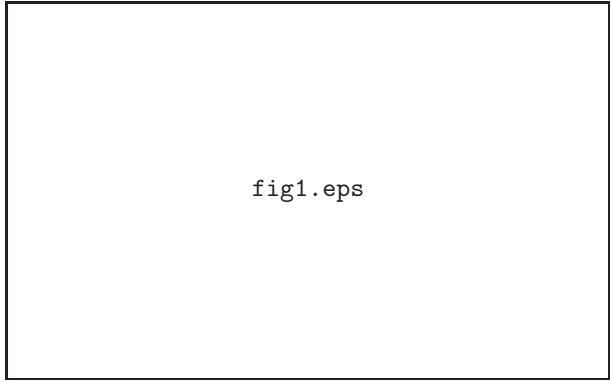

  \begin{center}
    \FigureFile(80mm,50mm){fig1.eps}
  \end{center}
  \caption{ Source distribution in the galactic coordinates. Filled and unfilled circles
  indicate the SiO detections and nondetections.
}\label{fig: l-b map}
\end{figure}

In summary, we observed 396 LAVs over three years. The SiO detections and
nondetections are summarized in tables 2 and 3, respectively. The spectra of
the SiO detections are shown in Figure 2a--2k.  In the \citet{gla01}'s list,
some objects are located very close to each other and they fall within one
beam diameter of the 45-m telescope (40$''$).  We observed only one object
of each such close pair, because they are unresolvable in our beam. We list
such objects lying closer than 12$''$ from each other in table 4, together
with those that have more than one name assignment in the LAV survey. The
objects in table 4 were thus considered to have been observed.  In the final
accounting of such overlapped observations etc., in 2003,  two sources were found to
be still missing; 4--22 and 11--307 (which was added in
the erratum) was not observed. Therefore we made an extra short observation
of these two stars in February 2004 together with a prvious marginal detection of 4--33. 
Thus all the LAVs listed by \cite{gla01} have been observed.

Because the surface number density of stars is so high in the galactic
center area, we often detected multiple emission peaks in a single beam.
When the velocity separation is more than 20 km s$^{-1}$, they are regarded
as different objects. In table 5, we listed such multiple detections in a
single beam (of HPBW=40$''$) and the possible assignment of the sources. The
associations with the OH/IR objects (\cite{sjo98}) are made in terms of the
radial velocity (within a few km s$^{-1}$) and the position (within $\sim
20''$).  The identified OH/IR objects were listed in table 6.  Furthermore,
identifications with previously known SiO/H$_2$O objects in this area of
the sky were also made. These are 1--72, 2--1, 2--18, and 3--266 which are
detected in \citet{miy01}, and 3--5, 3--6, 3--57, 3--88, 3--2885 in
\citet{deg02}; 3--3 was detected at 86.7 km/s in 1999 May, but not at this
time (2001 April).

When the velocity separation between multiple peaks is less than 20 km
s$^{-1}$, it is impossible to judge whether we are dealing with a single
object with multiple peaks, or two close-lying objects within the telescope
beam, unless a mapping observation has been made to separate the components.
For example, the SiO spectra of 5--157 in Figure 2d show peaks at $V_{\rm
lsr}=-2$ and 6 km s$^{-1}$. However, we found no significant sources within
30$''$ around this object. We tried to separate several multiple-peak
objects in the first year (for example, 9--8). However, because of a
shortage of observing time in the third year, we were unable to look further
into this issue. In table 5, we count 10 cases where source assignments
were impossible because of multiple peaks. In our statistics we took the
strongest peak to be the given object, for simplicity, and put an asterisk
sign after the name.



There is also further confusion in the SiO detections due to time variation.
For example, 13--18 (=10--3465 ) was observed twice in May 2001 and May 2002
, but emission was detected at different radial velocities, $-16$ and 38
km/s.  The star 13--4, is located 12.7$''$ away from 13--18, but was
not detected on May 2003. However, a careful check of the spectra in May
2003 gave weak enhancements of emission at 25 and 40 km/s.  We tentatively
assign the $-16$ km s$^{-1}$ component to be 13--18 and discard the May 2002
observations. It is highly possible that the $38$ km/s component is 13--4.





Several objects were detected in the off position during the position
switching sequence, at 7$'$ west of the first-on source (On-1) in right
ascension.  The off-position detections can easily be recognized; because of
the combination of the three objects with one off-position, all of the three
spectra have negative emission at the same radial velocity. These are
23--1198 (seen in the spectra of 6--151), and IRAS 17413$-$2909 (seen in the
spectra of 5--27). The $-60$ km/s component was detected in the spectra of
8--2 and 20--2631 (shown as off-1) . This off position was outside of the
observed area of \citet{gla01}, and was named to be SiO359.700+0.079
(17h44m35.9s, $-29^{\circ}09'01''$, J2000); no OH counterpart was found at
this position. This position is close to MSX5CG359.6949+00.0742 (17 h44
m36.12s $-29 ^{\circ}09 '27.4''$) with $F_{12}$=5.4 Jy. In addition, 24--108
was detected at the off position (in the spectra of 12--6) in April 2002. 
We observed this object in May 2003, but the emission was quite weak.


Because of the high chance of contamination by other objects in the
telescope beam, the positions and radial velocities of the previously known
OH sources were carefully checked to ensure that the cross-identifications
are correct.  For the sample of both OH and SiO detected objects, the
average radial-velocity difference between SiO and OH is
$<V_{\rm sio}-V_{\rm OH}>=$0.47 ($\pm 1.88$) km s$^{-1}$, 
where the standard deviation is shown in the parenthesis. The minimum and
maximum of the velocity differences are $-3.8$ and 4.3 km s$^{-1}$. These
values indicate that the identifications are almost certainly correct and
that no large systematic velocity shift occurs between SiO and OH.
In terms of OH and SiO velocity differences,  an interesting object,
13--200, was found to have an SiO radial velocity at $V^{\rm SiO}_{\rm
lrs}=23.7$ km s$^{-1}$. However, the position of this LAV is close to OH
0.178$-$0.055 with a separation of 5.1$''$, which has $V^{\rm OH}_{\rm
lrs}=-36.6$ km s$^{-1}$ (\cite{sjo98}). As noted in \citet{gla01}, the
separation of more than 5$''$ is too large to regard as an error in the
position measurement. The SiO detection with a large radial velocity
difference in this paper proves that these two are surely different stars.

\citet{mes02} surveyed the ISOGAL and MSX sources in the SiO $J=2$--1 $v=1$
transition; in their detections, 11 LAVs of \citet{gla01} were involved. We
checked our detection list and found nine detections (2 negative) in the SiO
$J=1$--0 $v=1$ and 2 transitions in the present paper.  Among their four
nondetections of LAVs, we got one detection (20--116) in the present paper.
The radial velocities of the SiO $J=2$--1 $v=1$ line are coincident with the
velocities of the SiO $J=1$--0 $v=1$ and 2 transitions within a few km
s$^{-1}$.

\citet{sjo02} made targeted surveys for 22 GHz H$_2$O and 43 GHz SiO
maser emission in the galactic center OH/IR stars using the Very Large
Array. Only 4 SiO detected objects overlap the LAVs in the
present sample, and these are marked $\natural$ in table 6. Radial velocities
also coincide very well with the velocities obtained in this paper. It is
curious that none of the objects in the present sample was detected in the
H$_2$O line. In general, the circumstellar H$_2$O maser line seems to be
weak in the LAVs in the galactic center (\cite{tay93}; \cite{lev95}).


\section{Discussion}

\subsection{Period--SiO Detection Rate}

In total, 180 LAVs from the original sample were detected in SiO 
with additional 16 objects. Figure 3 shows the period histogram of all
the observed LAVs and the SiO detection rate (line graph with filled
circles).  The line graph with unfilled circles in figure 3 indicates the OH
detection rate, which seems to correlate with the SiO detection rate quite
well.  Figure 3 clearly shows that the SiO detection rate increases with the
period.  The maximum detection rate occurs in the 600--700 d period bin.
Note that the average period of the \citet{gla01} sample is about 430 d.  A
slight decrease of the detection rate above $P>700$ d is probably a
statistical effect arising from the small numbers. The same graph was also
shown in \citet{ima02} for the sample of LAVs with the periods $P>400$ d 
and the period measurement quality $Q=3$ (Q=3 the highest quality)
preferentially observed in the first year of this program. 

In figure 3, we can recognize a small enhancement of the detection rate at
the short-period end. There are 16 sources with periods below 200 d; five
detections among these are 1--42, 4--28, 6--112, 7--9, and 14--38. The
period measurement quality of these LAVs are Q=2, 0, 1, 3, and 3,
where higher Q indicates better measurement quality. Therefore,
at least two LAVs of Q=0 or 1 are uncertain members of the $P=100$--200 d
bin. However, even so, we still have 3 firm detections in this bin.
Therefore, we get about a 20\% detection rate in this bin after this
correction, resulting a relatively flat, but non-zero, detection rate below
300 d. The significant SiO detection rate for periods below 200 d can also
be found in the database of stellar maser sources by \citet{ben90}, though
the listed sources are not at the same distances from the Sun.  In fact, a
significant number of semi-regular variables are known to be involved in the
short-period subsample in \citet{ben90}.

The SiO detection rate is approximately twice (or even triple for certain
period ranges) as high as for OH, indicating that the SiO maser survey has
doubled the number of stars with radial velocity data near the galactic
center. Note, however, that the SiO survey was a targetted survey and OH 
an unbiased (blind) survey. Therefore, the OH survey (\cite{sjo98}) detected
a number of objects which were not identified in NIR imaging;
they are simply not counted in the above comparison of the detection rate
between SiO and OH surveys. 

That the maser detection rate increases with the period of light variation
in LAVs has already been suggested by observations of stars near the Sun
(e.g., \cite{ben96}).  However, because of uncertainties in the distances of
these stars, the relation was not clearly demonstrated. In the present
sample, the distances to the LAV stars are almost equal and the correlation
with period is much more conclusive.

\subsection{Surface density}

The surface density distribution of the sample is useful for estimating an
approximate line-of-sight distribution of the LAVs. Figure 4 shows plots of
the cumulative number of objects, $N(r)$, within a given projected distance
from Sgr A*, $r$, normalized to 1 at the outermost projected distance. The
left panel shows the difference between the LAVs and the SiO-detected
objects (including all the detections), and the center panel shows the
difference between the $l^*>0$ and $l^*<0$ SiO detected objects. Here $l^*$
is the longitude offset from Sgr A*. The Kolmogorov-Smirnov test does
not give any significant probability of these two sets being statistically
different. The distributions can be well approximated by $N(r) \sim r^{1.3}
$. This function gives a surface density distribution for SiO or LAV objects
that varies as $\sigma (r) \sim r^{-0.7}$. If we assume that the source
distribution is spherically symmetric and with a power-law density profile, 
the number density of the sources
(per unit volume) is $\rho(R) \sim R^{-1.7}$, where R is the real distance
from the galactic center (Sgr A*) [see equation (2--43) of \citet{bin87}].
The present result fits quite well with the K-band surface density variation
near the Center ($\lesssim 6'$), $I_K \sim r^{-0.65}$, measured by
\citet{phi99}.

In this case, a simple computation gives that, for the sources within a projected
distance of 12$'$ ($\sim 30$ pc), 32\% of the objects are located between
the true radii 60 and 300 pc, and 12\% of the objects are between the radii 300 and
3 kpc. Furthermore, an integration shows that 31 \% of the observed objects in
the sample are actually within a radius of 30 pc.  The
contamination of the sample by SiO objects outside 30 pc is estimated to be
as follows; among the $\sim$200 SiO detections, 23 objects are outside 300
pc, 112 are sources between 30 and 300 pc, and 65 sources are within 30 pc.

However, the OH 1612 MHz source distribution looks slightly different as
shown in the right panel of figure 4 for 137 OH sources within 15.9$'$
[taken from \citet{sjo98}].  The OH integrated number distribution is
approximated by $N_{\rm OH}(r) \sim r^{0.8} $, leading to a steeper surface
density distribution  $\sigma (r) \sim r^{-1.2}$, and a number density
distribution $\rho(R) \sim R^{-2.2}$. In the right panel, we also plotted
the normalized integrated number (thin line) of LAVs with SiO in the
present sample (thick dots).  The Kolmogorov-Smirnov test gives the
probability of 2 \% that the two sets have the same statistical distributions.
Therefore, we conclude that the distributions {\it are} different. The
integrated number of LAVs with OH in the present sample is also plotted
as thin dots (67 objects). As expected, this line falls near the middle of
the two sets above. A statistical test does not give any conclusive
evidence that the distribution of the LAVs-with-OH set is different
from the above two sets (because of small numbers).


The surface density of our LAV sample increases toward the galactic center
more slowly than that for OH-sources. This is partly because the
near-infrared detection of LAVs is hindered by the high density of stars
near the galactic center.  In contrast, the OH surveys are less biased near
the galactic center. On the other hand, OH detection is enhanced towards the
long period end and, hence, toward massive objects [if the masses
(luminosities) and periods are correlated as normally assumed].  This fact
might imply a mass segregation effect for stars in the galactic center.
However, the statistical test for the present SiO sample with OH emission
gives a probability of greater than 20 \% that the two SiO subsets with and without OH have 
the same spatial distribution. Therefore, the conjecture of mass segregation
cannot be proved by using the present subsample.

Note that the power law obtained for SiO density distribution is close to
$R^{-1.75}$. If the mass density of the galactic center area also follows
the same rule, the rotational velocity as a function of radius
increases as $V(R)\sim R^{0.125}$. Because the $R^{-1.75}$ law makes various physical
quantities easy to estimate, we will assume that the density distribution
follows this law near the galactic center.




\subsection{Velocity Distribution}

Figure 5 shows a longitude--velocity diagram for all the SiO detected
sources. Here, $l^*$ is the galactic longitude offset from Sgr A* (the
dynamical center of the Galaxy; R.A.=$18^{\rm h}45^{\rm m}40^{\rm s}.05$,
Dec.=$-29^{\circ}00'27''.9$, J2000; \cite{rog94}).  Most of the radial
velocities fall within $\pm$200 km s$^{-1}$. Five extreme sources at
$|V_{\rm lsr}|>200$ km s$^{-1}$ occupy outlying positions. These sources
might be outer bulge objects in highly eccentric orbits, seen by chance in
the line of sight toward the galactic center [see discussion in
\citet{van92}].  These high-velocity objects appear almost evenly at positive and negative
velocities in the present sample.  However,  they appears only at the negative longitude 
side of the galactic center ($l^* <0$).

The regression-line analysis for the radial velocities gives $V_{\rm
lsr}=-6.1(\pm 5.6)+190.1(\pm 51.2) (l^*/$deg) km s$^{-1}$ and the standard
deviation from this line is 78.0 km s$^{-1}$ for all the 195
sources. If we remove the five extreme sources with
$|V_{\rm lsr}|>200$ km s$^{-1}$, the best fit gives $V_{\rm lsr}=-5.4(\pm
4.5)+209.8(\pm 41.9) (l^*/$deg) km s$^{-1}$. The best-fit slope, 190.1 km
s$^{-1}$ per degree, is compatible with the value, $\sim$190 km s$^{-1}$ per
degree, which was computed from the OH 1612 MHz data within 0.5 degree from
the galactic center (\cite{sjo98}).  Considering the slow rotation of the
inner Bulge out to $3^{\circ}$ of the galactic center,
$\sim$20 km s$^{-1}$ per degree (\cite{deg00}), we conclude that the
nuclear disk is rotating more rapidly than the inner Bulge.
The slight shift of the average velocity to the negative side may be
interpreted as contamination by foreground objects in stream
of stars forming part of a bar-like Bulge (\cite{izu95}; \cite{deg00}).

The overall structure of the SiO $l$--$v$ diagram is quite similar to the OH
$l$--$v$ diagram in the same region (see figure 3 of \cite{sjo98}). We can
recognize a hole in the SiO distribution at ($l^*=0.06^{\circ}$, $V_{\rm
lsr}=$ 60 km s$^{-1}$) in figure 5. The same hole can be seen in figure 3 of
\citet{sjo98}, though another hole at ($l^*=-0.04^{\circ}$, $V_{\rm lsr}=-20$ km
s$^{-1}$) in figure 3 of \citet{sjo98} does not exist in figure 5. The
position of this hole in the $l$--$v$ diagram corresponds roughly to the
dense CO and CS clouds at $V_{\rm lsr}\sim 75$ km s$^{-1}$, which can be
seen on the overlaid map of figure 5.  It is clear that the hole does not
simply arise from a bias in the distribution of the LAVs, arising from the
near-infrared sampling because it also appears in the unbiased OH survey
(\cite{sjo98}).

Another interesting characteristic of the SiO $l$--$v$ diagram is no strong
association of the SiO radial velocities with the ``240 pc molecular-ring"
features at $V_{\rm lsr}\sim -130$ -- $-100$ and $\sim 170$ -- 190 km
s$^{-1}$, which can be seen on the overlaid CO map in the left panel of
figure 5. This molecular-ring feature is extended in the area of $l=\pm
2^{\circ}$ and $b=\pm 0.5^{\circ}$, and is considered to be a manifestation of
the bar potential of the Bulge (\cite{bin91}).  On the line of sight toward
the galactic center, this ring is located at about 240 pc from the Center
itself.  A closer look at figure 5 shows that two objects at $V_{\rm
lsr}\sim170$ km s$^{-1}$ and $l^*=-0.12^{\circ}$, are on the
positive-velocity ring feature, and one is at
$V_{\rm lsr}\sim-140$ km s$^{-1}$ and $l^*=-0.17^{\circ}$;
these are 22--21.2 [($l$, $b$)=($-$0.170, 0.047)], 22--21.3, 
and 22--166 [($l$, $b$)=($-$0.229, 0.044)]. 
However, at $b=-0.4^{\circ}$ ($\sim 2'$),  the negative ring feature appears 
at $V_{\rm lsr}\sim -75$  km s$^{-1}$ in \citet{oka98}'s map.
Moreover, the positive-velocity ring feature actually appears very weakly 
at $V_{\rm lsr}\sim165$ km s$^{-1}$ and $l=-0.2^{\circ}$. The LAVs are
therefore not likely to be associated with the ring feature.
 
It is further reasonable that no parallelogram feature appears in the SiO and
OH $l$--$v$ diagram because the orbits cannot be cusped for the case of stars
in a bar-like Bulge (unlike the  gas; \cite{bin91}). Rather, the velocity
spread in the SiO $l$--$v$ diagram is much larger than the velocity spread of
the CO gas.  These facts indicate that random motions are considerably dominant for
stars in this region.


\subsection{The Mass of the Galactic Center}

From the velocity data of maser stars, we can obtain the gravitational mass
in the galactic center area in principle.  However, this is rather a
difficult task because only the line-of-sight velocities and projected
distances are known. In the past (\cite{sel87}; \cite{lin92a}), three
methods have been used for this purpose: the pressure-balance,
projected-mass, and virial methods, where we take the `pressure balance
method' to be the moment equation of the Boltzmann equation in the radial
direction, as given in \citet{sel87}.  Although none of these approaches are
really valid for application to the galactic center velocities, we first
used them to compute the gravitational masses in order to show that the
present sample gives results which are compatible with previous
calculations. The results from the pressure-balance and projected-mass
methods are presented as filled thin triangles and diamonds in figure 6.
For comparison, the previous OH 1612 MHz results (\cite{lin92a}) are also
plotted as unfilled thin triangles and diamonds in figure 6.


Because all three of the above noted methods rely on assumptions which are
not verified in the case of the galactic center sample in a strict (even
rough) sense, they are believed to yield errors of a factor of a few in the
computed masses.  For example, the pressure balance method uses an equation
derived in the radial direction and makes use of line-of-sight quantities.

In order to remedy the situation, we have developed a new method in this
paper. We first assume that the mass distribution is spherically symmetric.
We also assume that the enclosed mass within the radius $R$ is written by
\begin{equation}
M_R= m_1 +  m_2 R^{1.25}. 
\end{equation}
where $m_1$ corresponds to the galactic center black hole, and the second
term corresponds to the observed mass density distribution $\rho \propto
R^{-1.75}$ (and the surface density mass distribution $\sigma(r) \propto
r^{-0.75}$). In what follows, we will use cylindrical coordinates, $r$,
$\phi$, and $z$, where the $z$ axis is taken along the line of sight to the
galactic center (see figure 9).  In this coordinate system, $r$ is called the
projected distance from the galactic center. The observed quantity, the
average of $v_z^2$ between projected distances $r_n$ and $r_{n+1}$, can be
expressed using the Boltzmann one-particle distribution function, $f(r,\phi,z,v_r,v_{\phi}, v_z)$, as
\begin{eqnarray}
\lefteqn{ [\sum_i^{r_n<r_i<r_{n+1}} (v_{z,i})^2 \; ]/N_n }  \nonumber  \\
 \simeq & \int_{r_n}^{r_{n+1}} \int [ (v_z)^2 f] \ dz r d\phi dv_r dv_z dv_{\phi} dr \;/
[\int_{r_n}^{r_{n+1}}  \int  f \ dz r d\phi dv_r dv_z dv_{\phi} dr] 
\end{eqnarray} 
where $N_n$ is the number of the observed stars between  $r_n$ and
$r_{n+1}$, and the subscript $i$ indicates the i-th star in the sample. All
the quantities on the right side of the equation are regarded as continuos
functions of the coordinates, $r$, $\phi$, $z$, $v_r$, $v_z$, and $v_{\phi}$.
A partial integration with respect to $z$ in the numerator gives
\begin{equation}
\int_{r_n}^{r_{n+1}}  \int [z (v_z)^2 f]_{z=-\infty}^{z=+\infty} r d\phi dv_r dv_z dv_{\phi} dr -
\int_{r_n}^{r_{n+1}}  \int  [z v_z  (v_z \partial f/\partial z)] dz r d\phi dv_r dv_z dv_{\phi} dr .
\end{equation} 
Here the first term vanishes because the function $f$ tends to zero rapidly
as $z$ goes to infinity.  Using the collisionless steady-state Boltzmann
equation, the second integral can be related to the term $z v_z (F_z
\partial f /\partial v_z) $. Details are given in the Appendix. Here the $z$
component of the gravitational force $F_z$ is given as
$ F_z = -G M_R  z/ R^3 $,
where G is the gravitational constant and $R=(z^2+r^2)^{0.5}$. By
integrating this equation over the line of sight and all the other
coordinates with a density distribution
\begin{equation}
\rho(R) = \rho_0 \; R^{-1.75},
\end{equation} 
and assuming a symmetric velocity distribution, 
we obtain the equation to be solved:
\begin{equation}
 m_1 A_n +m_2 B_n =\ C_n,
\end{equation} 
where
\begin{equation}
 A_n = 1.0245 \ G \ ( r_{n+1}^{1/4}-r_n^{1/4})/(r_{n+1}^{5/4}-r_n^{5/4}) ,
\end{equation} 
and
\begin{equation}
 B_n =  0.7556 \ G \ ( r_{n+1}^{3/2}-r_n^{3/2})/(r_{n+1}^{5/4}-r_n^{5/4}) ,
\end{equation}
\begin{equation}
 C_n = (\sum_i^{r_n<r_i<r_{n+1}} v_{z,i}^2 \ ) / N_n . 
\end{equation}
The unknowns, $m_1$ and $m_2$, are solved by least squares with the
set of $ A_n$, $B_n$, and $C_n$ ($n=1,2,..n_e$), where $n_e$ is the number of divided regions 
in the cylinder. 

We calculated $A_n$, $B_n$, and $C_n$ ($n=1$ to 10) and solved $m_1$ and
$m_2$ by least squares. In the present line-of-sight velocity sample of 199
SiO maser stars within 30 pc from the galactic center. In order
to obtain accurate masses very near the galactic center, we added 10
SiO objects within 2$'$ of the galactic center, as found by
\citet{deg02}.  Furthermore, we excluded 5 extremely high velocity objects with
$|v_{\rm lsr}|>200$ km s$^{-1}$ from the fit. 
we divided the cylinder into 10 regions, where each has about 20 data points 
except the innermost cylinder which contains 10 points. The value of $m_1$
is somewhat sensitive to the velocity dispersion of the innermost 10 objects.
In the case with no rotational
motion ($V_{\rm Rot}=0$), we obtain
\begin{equation}
M_R= (2.69 [\pm 1.30]   + 0.53 [\pm 0.07] (R/pc)^{1.25})\times 10^{6} \; {\rm M}_{\odot}.
\end{equation}

We can also involve the average rotational motion. Details are given in
Appendix. Here we assume that the rotational velocity slowly increases with radius
 as $\sim r^{0.125}$. The average rotation speed can be obtained from the
observational quantities and we obtain the rotational speed at 30 pc as 
$V_0=86.2$ km s$^{-1}$. 
The mass distribution involving the rotation becomes
\begin{equation}
M_R= (2.69 [\pm 1.30] + 0.88 [\pm 0.07] (R/pc)^{1.25})\times 10^{6} \; {\rm M}_{\odot}.
\end{equation}

The thin and thick solid curves in figure 6 show the mass distributions for
the non-rotating and rotating cases, respectively. These masses are slightly
less than the values obtained using the other methods mentioned. Because of
the self-consistency of this method, which uses only the line-of-sight
velocities and the projected radii, we believe that the masses obtained in
this paper are more accurate than those found previously. Also the
effects of the foreground and background of the galactic center along the
line of sight are correctly taken into account.

\subsection{High Velocity Stars}

We can clearly see five distinct high velocity stars with $|V_{\rm lsr}|>200$
km s$^{-1}$ in figure 5. As noted before, these sources are located only on
the $l^*<0$ side, which is somewhat difficult to explain. However, because
of their small numbers, we can always argue that their presence is
accidental.

To understand the high velocity objects in greater detail, we consider the
energetics of stars in the galactic center. The energy of a particle moving
in the {\it fixed} gravitational potential can be written as
\begin{equation}
E/m =(1/2) v ^2 + U(R), 
\end{equation}
where m and $U(R)$ are the mass of the particle and the gravitational
potential at the radius, $R$, from the galactic center, respectively.  Here,
for convenience, we replace the term, $E/m$, by $U(R_{0})$, where $R_0$ is the
outermost radius to which the star can reach {\it if the orbit is linear}. 
Because the projected quantities,
$v_z^2$ and $r$, are always smaller than $v^2$ and $R$, respectively, we
have the inequalities
\begin{equation}
   R \geq r ,
\end{equation}
and
\begin{equation}
 2 G^{-1} R [U(R_0)-U(R)] = G^{-1}R v^2 \geq G^{-1}r v_z^2,
\end{equation}
where G is the gravitational constant. Here, the quantity, $G^{-1} r v_z^2 $
is often called the projected mass ($\equiv q$; \cite{bah81}). From the
mass distribution of the galactic center (equation 1), we can compute the
gravitational potential as
\begin{equation}
 U(R)= G ( -m_1 R^{-1} + 4 m_2 R^{0.25}) 
\end{equation}
Note that U(R) is a monotonically increasing function of $R$.

Figure 7 is a plot of the mass distribution, $M_R$, and the quantity $2
G^{-1} R [U(R_0)-U(R)]$, as a function of the real distance, $R$, for
various different $R_0$ (indicated as the value $R_0$).  We also plotted the
projected mass $q$ against projected radius $r$ for the SiO-detected objects
in figure 7 using the same axes.  Because of the inequalities (12) and (13),
the position of a star, ($R$, $G^{-1} R v^2$), must fall
to the upper right part of the point ($r$,  $G^{-1} r v_z^2$) 
in figure 7. 
More precisely, it will be above the line connecting the origin and 
the point ($r$,  $G^{-1} r v_z^2$) and at the right part of the line $R=r$.
In another words, the observed location of the projected quantities, (r, q),
gives the lower bound of (R, $2 G^{-1} R [U(R_0)-U(R)]$), in this diagram. 

Note that a particle in  a pure circular orbit in an edge-on plane (as
viewed from the sun) falls on the curve $M_R$ only when it moves along the
line of sight. In general, the orbit of a particle is not edge-on, and the
velocity vector is not along the line of sight. Therefore, the projected
mass of a particle falls on a much lower part of the $M_R$ curve. In
summary, the particles at the upper-right part of the diagram in figure 7
have higher energy in general. But note that the particles with high $U(R)$
also fall towards the lower part of the diagram due to a projection effect;
at best when $v_z=v$, the point, ($R$, $G^{-1} R v^2$), moves on a straight
line through the origin.

With the help of the projected mass--radius diagram of figure 7, we can make
subsamples of the set and investigate the statistical properties of the high
$U(R_0)$ subsample. Let us divide the sample by the $q=2 G^{-1} r
[U(30$pc$)-U(r)$)] curve and call the upper subset ($q>2 G^{-1} r
(U(30$pc)$-U(r)$) as a high $q$ subset.  Note that this subset consist of
the objects distributed over all the projected distances, but weighted more
to the largest values.  We tried to find statistical differences in
$K$-magnitude and period between the high and low $q$ subsets. However, no
strong effects were found. If the stellar system is in equipartition of
energy (though this is somewhat absurd in a collisionless system), the light
mass particles may have larger specific energy $E/m$. However, we do not
find any period or $K$-magnitude segregation effect on energy. Both
luminosity and period are roughly increasing functions of stellar mass.
Therefore, our finding, i.e., that the statistical properties of the two
sets are similar, indicates that there is no strong mass segregation effect.
Of course, the low-$q$ subset is not purely one whose members
have low energies but, in fact, comprises a mixture of stars
with low and high $U(R_0)$ because of the projection effect. Therefore, it
is understandable that they do not exhibit any strong differences in $K$ and
period statistics.

We can also divide the high $q$ set into two; $V_{\rm lsr}>0$ and $V_{\rm
lsr}<0$.  This is because the SiO maser sample may exhibit a streaming
motion of stars along the Bulge bar. \citet{izu95} demonstrated that the
Bulge objects in front of the galactic center tend to exhibit a negative
radial velocity and the objects behind a positive velocity, due to the
streaming motions of stars in the bar-like Bulge.  The same tendency was
also observed in the sample of SiO maser stars within 3 degrees of the
galactic center (\cite{deg00}).  This front and back effect may produce a
systematic difference in the $K$ magnitudes of the stars (\cite{deg02}); the
frontward stars must be systematically brighter than the background ones due
to distance or absorption by circumnuclear dust. Because the high-$q$ objects
are considered to be located in general (but not strictly always) outside
$R$=30 pc, circumnuclear absorption is thought to have a significant
influence on the $K$ magnitudes.

Figure 8 shows a plot of $K$ magnitude against period for the LAVs. The
filled and unfilled circles indicate the objects with positive and negative
$V_{\rm lsr}$ both with $q>2 G^{-1} r [U(30$pc$)-U(r)$]. 
Apparently, figure 8 indicates that the positive-$V_{\rm lsr}$ objects are
fainter at $K$ than the negative ones.  The average
$K$ magnitudes are 9.44 ($\pm 1.21$) and 8.96 ($\pm 0.79$) for the 25 positive-
and 25 negative-velocity sets respectively. The t- and F-tests give a
probability of 11 \% for the average $K$ magnitude difference and a
probability of 4 \% for the standard-deviation difference being produced by
the same distribution function. In other words, with more than 89\%
probability, two sets are statistically different. Therefore, if this
average-magnitude difference is interpreted as being produced due to 
absorption by circumnuclear dust, the negative-velocity stars are in front
of the galactic center and the positive-velocity stars are behind it.

We can recognize in figure 8 a slight shift in distribution of the two sets
with respect to period. However, neither the t- nor the F-test gives any
probability higher than 78 \% for an average-period difference or a
standard-deviation difference between the two sets. Therefore, we regard
the period distributions in the two sets as being the same.

The systematic tendency towards brighter $K$-magnitude in the
negative-velocity subset appears more strongly in the smaller subsets.  For
the top five high-velocity stars with $|V_{\rm lsr}|>190$ km s$^{-1}$, we
obtain $K_{ave}$=9.25 ($\pm 0.71$) and $K_{ave}$=8.23 ($\pm 0.30$) for two
positive and three negative high-velocity LAV stars. For the top 23 high-$q$
stars, we obtain $K_{ave}$=9.63 ($\pm 0.99$) and $K_{ave}$=8.93 ($\pm 0.94$)
for 11 positive and 12 negative high-velocity stars. Here the numbers in
parentheses are standard deviations. Though these results are not
statistically significant because of small samples, we believe that the
tendency appears more drastically in the extremely high-velocity stars.

The above finding, i.e., that the negative-velocity stars are located
relatively in front of and the positive-velocity stars behind the galactic
center, indicates that they share the same dynamical property
 as the stars in the Bulge (\cite{deg01}).\footnote{
It suggests that orbital motions of these stars are close to those 
of the $x_1$ orbit family in a bar potential (\cite{bin98}). 
Because the $x_2$ orbits are less elongated 
and have smaller orbital velocities than the $x_1$ orbits, 
the present high-velocity star subsample does not seem to
show properties of the $x_2$ orbit family.}
 However, this fact does not immediately suggest
that all of the high-velocity stars belong to the Bulge population and that their
real distances from the galactic center are much farther than the projected
distance, say about 30 pc.  The average difference of 0.4 mag
corresponds to a distance difference of about 1.5 kpc, if the magnitude
difference is produced purely by distance effects.  On the other hand, the
distance difference is computed to be about 120 pc, when the extinction
model of $A(K) \sim 3$ kpc$^{-1}$ is applied near the galactic center.

Though interstellar extinction by dust grains must have a strong influence
on the $K$-magnitudes obtained, it is hard to quantify it because the dust
distribution is irregular on scales of 1--5$'$ towards the galactic center. 
Because the present high-$q$ objects are not concentrated in a particular
region in the $24'\times 24'$ area, it is unthinkable that the above noted
difference in the $K$-magnitudes is created only by the irregular
distribution of the dust toward the galactic center.

The other notable point on the extremely high velocity stars is the
lopsidedness of their distribution; all of the six extremely high-velocity
stars with $|V_{\rm lsr}|>190$ km s$^{-1}$ appear in the $l^*<0$ side of the
sky plane in the present sample.  If we assume that an object should appear
equally in both half-planes, the probability for all six to lie
on one side of the plane is $1/32\sim 3$ \%.  Of course, we can always argue
that it is due to statistical fluctuations.  In fact, we find an OH
high-velocity source, OH 0.335$-$0.180 with $V_{\rm lsr} =-355.1$ km
s$^{-1}$, on the $l^*>0$ side of the plane in \citet{sjo98}, outside 
the present survey area.
 
The distribution of the circumnuclear molecular ring (\cite{wri01}) is very
asymmetric about Sgr A*, and the MSX 21 $\mu$m map in the same region shows
heavy concentrations of bright objects on the $ l^*>0$ side
(\cite{pri01}).  Considering the dynamical (rotation) time scale of about 2
My, and the lifetime of molecular clouds of about 1--10 My, the
lopsidedness of the interstellar matter can be regarded as a relatively
short-lived phenomenon. The ages of AGB stars are between 30 My (about 9
$M_{\odot}$) and 2 Gy (about 2 $M_{\odot}$) (\cite{vas93}; \cite{mou02}).
Therefore, such a lopsided distribution of the high velocity stars (if it is
not a statistical fluctuation) may be more or less a reflection of an
asymmetry of the gravitational potential.  \citet{kim03} made a numerical
simulation of the dynamical friction of stars in the central cluster within
30 pc and found that massive stars in a cluster can sink to the 
center within a relatively short time.  By a counter-reaction, low-mass
stars can be ejected from the cluster and can be observable as high-velocity
stars. Though it is related to the disruption of young stellar clusters, the
possible lopsided distribution of aged AGB high-velocity stars might be
explained by a chance encounter of these AGB stars with such young clusters.
However, we have not yet been able to reach a unified view of the high-velocity
star phenomenon in the galactic center from the present data.

\section{Conclusion}

We have surveyed $\sim 400$ large-amplitude variables within a $24'\times 24'$
square about the galactic center, and obtained 180 detections (with additional
16 detections other than LAVs) in the SiO maser lines. The SiO detection
rate of $\sim$48\% is comparable to that in previous SiO surveys of
color-selected Bulge IRAS sources. The SiO detection rate increases with the
period of light variation, and is well correlated with the OH detection
rate.  The longitude-velocity diagram of the SiO sources has been revealed
to be quite similar to the OH $l$--$v$ diagram. These facts suggest that
the large-amplitude variables in the galactic nuclear disk are mass-losing
stars in the AGB phase, quite similar to the IRAS sources in the inner
Galactic bulge.

We also analyzed the SiO radial velocity data and obtained the mass
distribution of the galactic center area. The mass of the central black hole
that we have deduced, $2.7 (\pm1.3) \times 10^6 M_{\odot}$, and the mass
within 30 pc, $6.5 (\pm 0.7) \times 10^7 M_{\odot}$, are 
 more accurate than the previous estimates for these
quantities. From analysis of the projected mass vs radius diagram, we found
a tendency among the high-velocity sources that
the subset with negative line-of-sight velocity is systematically brighter
than the subset with positive line-of-sight velocity. This results from the
fact that the the subsample with negative line-of-sight velocity is in front
of the galactic center and the subsample with positive velocity is behind
it. This tendency, which also applies to the Bulge SiO maser sources,
strongly suggest the presence of streaming motion in the present
nuclear-disk LAV sample.

The authors thank Dr. A. Winnberg for the useful comments.
One of authors (I.S.G.) thanks the National Astronomical Observatory for
providing him with a visiting fellowship for this work. This research was
made use of the SIMBAD database operated at CDS, Strasbourg, France. It
was partly supported by Scientific Research Grant (C2) 12640243 of
the Japan Society for Promotion of Sciences.

\section*{Appendix. Theory for Computing the Enclosed Mass}

We assume that the mass distribution is spherically symmetric. We also
assume that the enclosed mass within the radius $R$ is given by
\begin{equation}
M_R= m_1 +  m_2 R^{1.25}, 
\end{equation}
where the first term corresponds to the mass of the  central black hole, and
the second term corresponds to the general mass density distribution
(including dark matter) $\rho_R \sim R^{-1.75} $
(and the surface density mass distribution $\sigma(r) \sim r^{-0.75}$). Here,
we adopt cylindrical coordinates, $r$, $\phi$, and $z$, where the $z$ axis
is taken along the line of sight to the galactic center, $r$ is the
projected distance from the galactic center, and the direction of $\phi=0$
to the galactic plane (see figure 9). The collisionless Boltzmann equation
in a steady state can be written in cylindrical coordinates [for example,
\citet{bin87}]
\begin{equation}
v_r\partial f/\partial r + (v_{\phi}/r)\partial f/\partial \phi + v_z\partial f/\partial z
+(v_{\phi}^2/r + F_r)\partial f/\partial v_r  - r^{-1} v_r v_{\phi}\partial f/\partial v_{\phi} 
+F_z \partial f/\partial v_z \ =0,
\end{equation}
where the $\phi$ component of the gravitational force $F_{\phi}$ vanishes 
because of the spherical symmetry.
The average of $v_z^2$ in the area between $r_n$ and $r_{n+1}$ 
can be expressed using the Boltzmann one-particle distribution function, $f$, as
\begin{eqnarray}
\lefteqn{ [\sum_i^{r_n<r_i<r_{n+1}} (v_{zi})^2]/N_n \ }  \nonumber \\ 
\simeq & \int_{r_n}^{r_{n+1}} \int [ (v_z)^2 f] \ dz r d\phi dv_r dv_z dv_{\phi} dr/
[\int_{r_n}^{r_{n+1}}  \int  f \ dz r d\phi dv_r dv_z dv_{\phi} dr] ,
\end{eqnarray} 
where $N_n$ is the number of stars observed between  $r_n$ and $r_{n+1}$,
and the denominator is for normalization. All the quantities on the right
side of the equation are regarded as continuos functions of the coordinates,
$r$, $\phi$, $z$, $v_r$, $v_{\phi}$, and $v_z$. The partial integral with
respect to $z$ can convert the numerator to
\begin{equation}
\int_{r_n}^{r_{n+1}}  \int [z (v_z)^2 f]_{z=-\infty}^{z=+\infty} r d\phi dv_r dv_z dv_{\phi} dr -
\int_{r_n}^{r_{n+1}}  \int  [z v_z  (v_z \partial f/\partial z)]  dz r d\phi dv_r dv_z dv_{\phi} dr,
\end{equation} 
Here the first integral vanishes because the distribution function $f$ tends
rapidly to zero at infinity.  Using the collisionless steady-state Boltzmann
equation (16), the second integral can be related to the term $z v_z (F_z
\partial f /\partial v_z) $.
Here  the $z$ component of the gravitational force $F_z$ is given as
$ F_z = -G M_R  z/ R^3 $
where $R=(z^2+r^2)^{0.5}$. We perform the integral (18) over all the
coordinates with a density distribution
\begin{equation}
\rho_R = \rho_0 \; R^{-1.75}.
\end{equation} 
Note that
\begin{equation}
\int_{-\infty}^{+\infty} f  dv_r dv_z dv_{\phi} = \rho_R .
\end{equation}
When the system is not rotating, further partial integral in expression (18)
after the above noted replacement easily results in equation (5).

We have to involve the average rotational motion by the replacing
\begin{equation}
 v_r =  v_{0,r} + V_{\rm rot,r}
\end{equation}
\begin{equation}
 v_{\phi} =  v_{0,\phi} + V_{\rm rot, \phi}
\end{equation}
and
\begin{equation}
 v_z =  v_{0,z} + V_{\rm rot,z}
\end{equation}
where $V_{\rm rot,r}$, $V_{\rm rot, \phi}$, and $V_{\rm rot,z}$ are the  $r$, $\phi$, and
$z$, components of the rotational velocity. Here we assume the rotational
axis is perpendicular to the galactic plane (i.e., perpendicular to z axis). 
We assume the rotational motion is fastest at equator, and gets slower
according to a cosine law when approaching the pole (see figure 9).  In
such a case, the rotational velocity components can be written using the
rotational velocity at the equator, $V_{\rm Rot}$, as
\begin{equation}
 V_{\rm rot,r}= -V_{\rm Rot} \; z \, cos(\phi)/R, 
\end{equation}
\begin{equation}
 V_{\rm rot, \phi}=V_{\rm Rot} \; z \, sin(\phi)/R,
\end{equation}
and
\begin{equation}
 V_{\rm rot,z}=V_{\rm Rot}\; r \, cos(\phi)/R
\end{equation}
(see figure 9).
Here we assume that the rotational velocity $V_{\rm Rot}$ at the equator is
given by the law corresponding to $\rho \ \sim R^{-1.75}$,
\begin{equation}
 V_{\rm Rot} \ = \ V_{\rm 0} (R/r_0)^{0.125} ,
\end{equation}
where $V_0$ is the rotational velocity at the equator at $r_0=$ 30 pc.  
A more realistic case of $V_{\rm Rot}$ as a
function of radius can also be calculated if the integral converges, but we
do not pursue such a complex case here.

We perform integral (18)  with respect to $z$, $r$, $\phi$, $v_{0,z}$,
$v_{0,r}$, and $v_{0,\phi}$, assuming that the distribution function is
symmetric in $z$, $v_{0,z}$, $v_{0,r}$, and $v_{0,\phi}$ [for example,
$f(r,\phi,z,v_{0,z},v_{0,r},v_{0,\phi})
=f(r,\phi,z,-v_{0,z},v_{0,r},v_{0,\phi})$]. 
In the partial integral, which must be made to convert the derivatives 
$\partial f/\partial r$, etc., to $f$,  
the odd terms of these variables vanish due to the symmetry. 
The only remaining terms come from the part, $z v_z F_z \partial
f/\partial v_z$ (leading the term $m_1 A_n + m_2 B_n$), and the part 
(only in the rotational case),
$ z v_z v_r \partial f /\partial r$  (leading the term $V_{0}^2\ D_n$).
We obtain the equation to be solved:
\begin{equation}
m_1 \ A_n +m_2 \ B_n  = C_n + V_0^2 \ D_n,
\end{equation} 
where 
where 
\begin{equation}
 A_n = 1.0245 G \ (r_{n+1}^{1/4}-r_n^{1/4}) /(r_{n+1}^{5/4}-r_n^{5/4}),
\end{equation} 
\begin{equation}
 B_n = 0.7556 G \ (r_{n+1}^{3/2}-r_n^{3/2}) /(r_{n+1}^{5/4}-r_n^{5/4}),
\end{equation}
\begin{equation}
 C_n = (\sum_i^{r_n<r_i<r_{n+1}} v_{z,i}^2 \ ) / N_n . 
\end{equation}
and
\begin{equation}
 D_n = 0.5667 G \ r_0^{-1/4} \ (r_{n+1}^{3/2}-r_n^{3/2}) /(r_{n+1}^{5/4}-r_n^{5/4}) ,
\end{equation}
The unknowns, $m_1$ and $m_2$,  are solved by least squares with
the set of $ A_n$, $B_n$, and $C_n$ ($n=1,2,..n_e$), where $n_e$ is the total number 
of divided regions in the cylinder. The rotational term $D_n$ appears due to
the centrifugal force.

The average rotational velocity,  $V_{\rm 0}$, can be computed from the following equation; 
\begin{equation}
\sum_i^{all} v_{zi} \ cos(\phi_i)]/N  \simeq \int v_{z} \ cos(\phi) f dr dz r d\phi dv_r dv_z dv_{\phi}
/ (\int f dr dz r d\phi dv_r dv_z dv_{\phi}),
\end{equation}
where N is the total number of the observed particles and the integration
must be made over all the coordinates. By integrating the right hand side
after replacing $v_z$ by equations (24) -- (27), we get
\begin{equation}
V_{\rm 0} = 3.726 [\sum_i^{all} v_{zi} \ cos(\phi_i)]/N.
\end{equation}

All the computations described above were made with help of 
computer-algebra software. A key aspect of the present analysis is that the
functional form of the density distribution is given so that we can
integrate the distribution function. The present analysis does not assume
any special velocity distribution function except symmetry with respect
to the velocities,
$v_{0,z}$, $v_{0,r}$, and $v_{0,\phi}$.
This is considered to be the same as an isotropic velocity distribution,
though it is somewhat obscured because of the cylindrical coordinates.


In this method, the effect of the objects to the front and rear along the
line of sight is correctly taken into account.  Because of the self
consistency, i.e., using only the line-of-sight velocities and projected
radius, we believe that the mass distributions we have obtained are more
accurate than those obtained with the projected-mass or pressure-balance
methods [for the reference, see \citet{bah81}].



%
%
\begin{table*}
  \caption{additional stars observed}\label{tab:table1}
\begin{tabular}{lrrrrrrrrrc}
  \hline\hline                           
Name & \multicolumn{6}{c}{R. A. (J2000)  Dec.} & $P$ & $K_{\rm av}$ & $A$ & comment \\ 
     & h & m & s & $\circ$ & $'$ & $''$  & (d) & (mag) & (mag) & \\
	 \hline
6-154  &17 &45 &48.6 &$-29$ &09 &55.8  & ---  & --- & --- & erratic possibly double \\ 
7-389  &17 &45 &59.7 &$-29$ &02 &43.4  & 362 & 10.5 & 1.5 & \\
23-255 &17 &45 &00.6 &$-28$ &59 &54.7  & 426 & 10.4 & 1.3 & \\
\hline
\end{tabular}
\end{table*}
\tabcolsep 1pt
\begin{longtable}{lrrrrccrrccc}
\multicolumn{12}{c}{Table 2. SiO detections}\\
\hline\hline
%
          &&& \multicolumn{4}{c}{SiO $J=1$--0 $v=1$} & \multicolumn{4}{c}{SiO $J=1$--0 $v=2$} & \\
name &$l$&$b$& $V_{\rm peak}$ & $T_{\rm peak}$ & Flux & RMS & $V_{\rm peak}$ & $T_{\rm peak}$ & Flux & RMS & Obs. Date\\
     &($^{\circ}$)&($^{\circ}$)& {\tiny (km s$^{-1}$)} & {\tiny (K)} & {\tiny (K km s$^{-1}$)} & {\tiny (K)} 
		    & {\tiny (km s$^{-1}$)} & {\tiny (K)} & {\tiny (K km s$^{-1}$)} & {\tiny (K)} & (yymmdd.d) \\
\hline
1--2     &0.089&0.007   &$-$32.3 & 0.093 & 0.074 & 0.034 &$-$30.5 & 0.185 & 0.087 & 0.035 &  020309.2\\
1--8     &0.095&$-$0.007  & 79.1 & 0.417 & 1.342 & 0.049 & 79.0 & 0.335 & 0.707 & 0.050 &  010220.3\\
1--31    &0.113&0.063   &$-$39.2 & 0.183 & 0.726 & 0.038 &$-$38.0 & 0.146 & 0.561 & 0.043 &  010311.2\\
1--42.1  &0.074&0.023   & 43.5 & 0.133 & 0.346 & 0.041 & 44.0 & 0.132 & 0.175 & 0.043 &  020525.1\\
\end{longtable}

\clearpage
\begin{longtable}{lrrrrcc}
  \caption{Negative results} 
  \multicolumn{7}{c}{Table 3. Negative results}\\
  \hline\hline  
GMCS Name & $l$ & $b$ & RMS($v=1$) & RMS($v=2$) & ObsDate1 & ObsDate2 \\
          & ($^{\circ}$) & ($^{\circ}$) & (K)     &  (K)     & (yymmdd.d) & (yymmdd.d) \\
  \hline
1--19       &0.122&0.049   & 0.041 & 0.040 & 020522.1 & 010525.1\\
1--26       &0.095&0.075   & 0.066 & 0.062 & 020224.3 & 010526.0\\
1--37       &0.074&0.015   & 0.047 & 0.047 & 020524.0 & 010522.0\\
1--41       &0.086&0.056   & 0.048 & 0.051 & 010526.0 & 020520.0 \\
1--46       &0.060&0.017   & 0.046 & 0.046 & 010221.2 & 020424.2\\
\end{longtable}
\begin{longtable}{l|llll}
  \caption{Objects not specially pointed, but nevertheless observed within the same telescope beam} 
  \multicolumn{5}{c}{Table 4. Objects not listed in Tables 2 and 3, but regarded as observed}\\
  \hline\hline
GMCS Name & Observed as & table  & separation & comments \\
          &             &listed  & ($''$)     &          \\
  \hline
  \multicolumn{4}{c}{the same sources} & \\
2--101   &  1--72   & 2 & 1.9 &  detected also on 20305.3 \\
5--159   & 20--2631 & 2 & 0.7 &  detected also on 10527.1 \\
9--75    & 12--46   & 3 & 2.1 &  nondetection \\ 
10--171  &  1--8    & 2 & 0.7 &  deteceted also on 20316.3 \\
\hline
\end{longtable}
%
\setcounter{table}{4}
\begin{longtable}{lclcl}
  \caption{Multiple detections} \label{tab:table5}
  \multicolumn{5}{c}{Table 5. Multiple detections and their assignment}\\
  \hline\hline
Observed as &  $V_{\rm lsr}$  & Assignment  & separation & comments \\
            &   (km s$^{-1}$) &             & ($''$)     &          \\
  \hline
1--42.1$^{*}$ & 43.8 & \multicolumn{2}{l}{assigment impossible}  &  1--182(17.8"; negative)\\
1--42.2       & 57.3 & \multicolumn{2}{l}{assigment impossible}  &  \\
2--11.1    & $-$23.0 & 2--11 & 0.0 &  \\ 
2--11.2    & $-$6.6    & $\sim$OH359.998$-$0.005 & 19.5 & this is not 2--10 \\
2--26.1    & $-$63.9 &  \multicolumn{2}{l}{assigment impossible}  &  2--11 (26.2$''$; $-23$ kms$^{-1}$)   \\
\end{longtable}

%
%
%
\begin{longtable}{l|rrrcl}
  \caption{Comparison with OH} \label{tab:table6}
  \multicolumn{6}{c}{Table 6. Comparison with OH observations}\\
  \hline\hline
Observed as &  $V^{SiO}_{lsr}$  & OH Name  & $V^{OH}_{lsr}$ & separation & comments \\
          &     (km s$^{-1}$)   &listed  &  (km s$^{-1}$)  &   ($''$)   &      \\
  \hline
1--1890 &   23.6 & 0.083+0.063$\natural$ & 24.8 & 1.7 &   \\
2--10 $\dag$  &   $-$6.9  & 359.998$-$0.005 & $-$6 & 15.8 & this is not 2--10 \\
2--11.2$\dag$ &   $-$6.6  & 359.998$-$0.005 & $-$6 & 19.5 & this is not 2--10 \\
2--28 & 36.0    & 0.037$-$0.003 & 34.9 & 2.9 &   \\
2--320 & $-$4.7   & 0.060$-$0.018 & $-$4.3 & 1.7 &   \\
\end{longtable}

\begin{figure}
\vspace{-1cm}
  \begin{center}
    \FigureFile(170mm,240mm){fig2a.eps}
  \end{center}
  \caption{a. Spectra of the SiO $J=1$--0 $v=1$ and 2 lines for the detected sources.}\label{fig:l-v.diagram}
\end{figure}
\setcounter{figure}{1}
\vspace{-1cm}
\begin{figure}
  \begin{center}
    \FigureFile(170mm,240mm){fig2b.eps}
  \end{center}
  \caption{b.} 
\end{figure}
\setcounter{figure}{1}
\begin{figure}
\vspace{-1cm}
  \begin{center}
    \FigureFile(170mm,240mm){fig2c.eps}
  \end{center}
  \caption{c.} 
\end{figure}
\setcounter{figure}{1}
\begin{figure}
\vspace{-1cm}
  \begin{center}
    \FigureFile(170mm,240mm){fig2d.eps}
  \end{center}
  \caption{d.} 
\end{figure}
\setcounter{figure}{1}
\begin{figure}
\vspace{-1cm}
  \begin{center}
    \FigureFile(170mm,240mm){fig2e.eps}
  \end{center}
  \caption{e.} 
\end{figure}
\setcounter{figure}{1}
\begin{figure}
\vspace{-1cm}
  \begin{center}
    \FigureFile(170mm,240mm){fig2f.eps}
  \end{center}
  \caption{f.} 
\end{figure}
\setcounter{figure}{1}
\begin{figure}
\vspace{-1cm}
  \begin{center}
    \FigureFile(170mm,240mm){fig2g.eps}
  \end{center}
  \caption{g.} 
\end{figure}
\setcounter{figure}{1}
\begin{figure}
\vspace{-1cm}
  \begin{center}
    \FigureFile(170mm,240mm){fig2h.eps}
  \end{center}
  \caption{h.} 
\end{figure}
\setcounter{figure}{1}
\begin{figure}
\vspace{-1cm}
  \begin{center}
    \FigureFile(170mm,240mm){fig2i.eps}
  \end{center}
  \caption{i.} 
\end{figure}
\setcounter{figure}{1}
\begin{figure}
\vspace{-1cm}
  \begin{center}
    \FigureFile(170mm,240mm){fig2j.eps}
  \end{center}
  \caption{j.} 
\end{figure}
\setcounter{figure}{1}
\begin{figure}
\vspace{-1cm}
  \begin{center}
    \FigureFile(170mm,240mm){fig2k.eps}
  \end{center}
  \caption{k.} 
\end{figure}
\begin{figure}
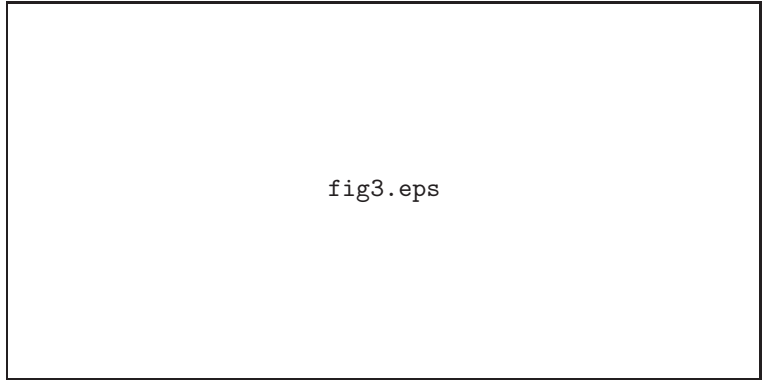

  \begin{center}
    \FigureFile(100mm,50mm){fig3.eps}
  \end{center}
  \caption{Histogram of period and detection probability (line graph). The 
  shaded and unshaded area of the histogram indicate the SiO detection and 
  nondetection, respectively. The line graphs with filled and unfiled circles 
  indicate the detection rate of SiO and OH masers, respectively, for the observed sources
  (unit at the right vertical axis).}\label{fig:histogram}
\end{figure}
\begin{figure}
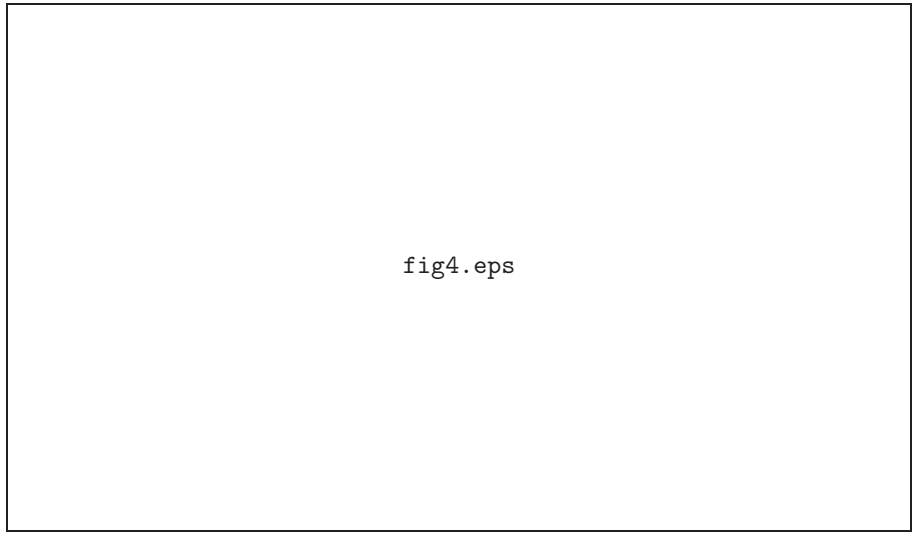

  \begin{center}
    \FigureFile(120mm,70mm){fig4.eps}
  \end{center}
  \caption{Normalized accumulation number with respect to $r^2$ 
  for all the LAVs and SiO detections (left panel), 
  that for  SiO detections at $l^*>0$ and at $l^*<0$ (center panel), 
  and that for OH sources in a OH survey, and LAV with OH, and LAV with SiO 
  in the present sample (right panel). The OH surface-density distribution  
  in the OH survey ($N\sim r^{0.8}$) is apparently different from the distribution 
  of the LAVs with SiO (and that of LAVs; $N\sim r^{1.3}$). The holizontal axis 
  is taken as $r^2$, so that the constant surface density 
  gives a straight line ($N\sim r^2$) in this figure.
  }\label{fig: Accm diagram}
\end{figure}
\begin{figure}
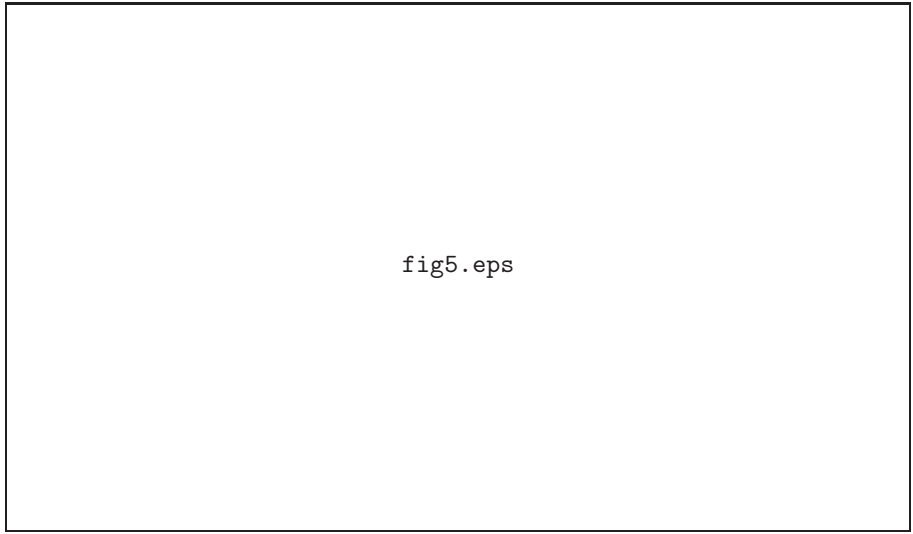

  \begin{center}
    \FigureFile(120mm,70mm){fig5.eps}
  \end{center}
  \caption{Longitude-Velocity diagram of the SiO detections. Filled circles indicate the objects with $r<5'$
  on the left panel and with $r<3'$ on the right panel. The regression lines are indicated by broken or solid lines 
  for each sets. The CO (left panel) and CS (right panel) $J=1$--0 $l$--$v$ diagram 
  at $b\sim -4'$ (\cite{oka98}; \cite{tsu99}) are overlaid.}\label{fig: l-v diagram}
\end{figure}
\begin{figure}
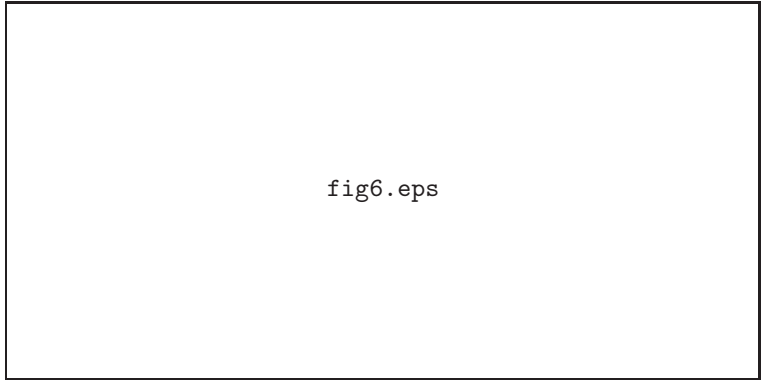

  \begin{center}
    \FigureFile(100mm,50mm){fig6.eps}
  \end{center}
  \caption{Mass distribution in the galactic center area. Filled triangles and diamonds indicate 
  the enclosed masses obtained for SiO maser data set
  by the pressure-balance (PB) and projected-mass (PM) methods, respectively. 
  Open triangles and  diamonds indicate the mass from the OH-maser data set 
  (\cite{lin92a}) with  the same two methods. The thick and thin curves with filled squares 
  and circles indicate the mass obtained by the present method for rotational 
  and non-rotational cases, respectively. statistical errors 
  are also  shown by tickmarks.  }\label{fig:M-R diagram}
\end{figure}
\begin{figure}
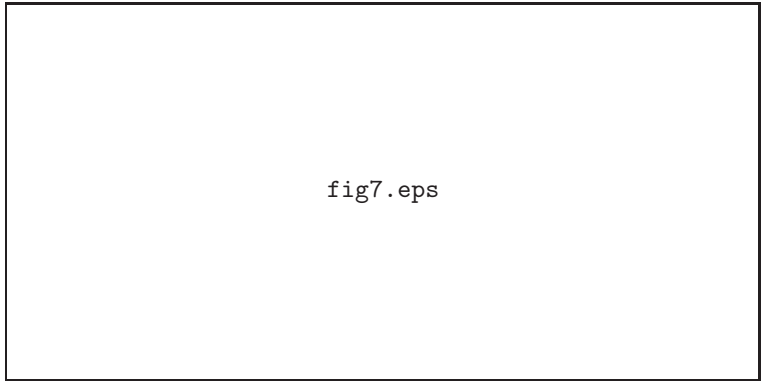

  \begin{center}
    \FigureFile(100mm,50mm){fig7.eps}
  \end{center}
  \caption{Plot of projected mass versus projected distance. Circles indicate
  the projected mass for indivisual object. 
  The dotted and broken curves indicate $q=2 G^{-1} R [U(R_0)-U(R)]$, where $R_0$ is indicated. 
  Thick solid curve with sign "$M_R$" indicate the calculated enclosed mass.  
  Note that these circles and curves are plotted with r and R, respectively, but are
  plotted using the same axis in this figure. Because of inequalities (12) and (13),
  the projected quantities, $(r, q)$, give the lower bound for 
  (R, $2 G^{-1} R [U(R_0)-U(R)])$ for the particular object.  
  Several extremely high velocity
  object are located out of this diagram and are not shown.}\label{fig:projected mass diagram}
\end{figure}
\begin{figure}
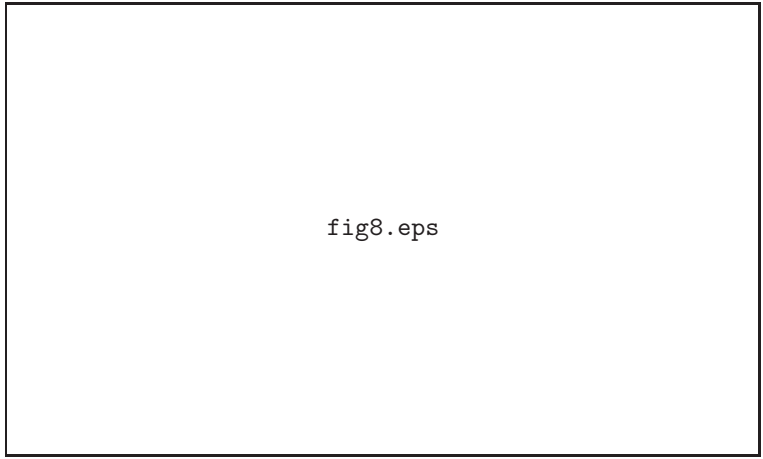

  \begin{center}
    \FigureFile(100mm,60mm){fig8.eps}
  \end{center}
  \caption{K Magnitude--Period diagram for SiO detected objects. Open circles indicate
 the stars with   $q>2G^{-1}r[U(30pc)-U(r)]$ and $V_{lsr}>0$,
  filled circles the stars with $q>2G^{-1}r[U(30pc)-U(r)]$ and $V_{\rm lsr}<0$,
  and the crosses the stars with $q<2G^{-1}r[U(30pc)-U(r)]$.
  }\label{fig: K-P diagram}
\end{figure}
\begin{figure}
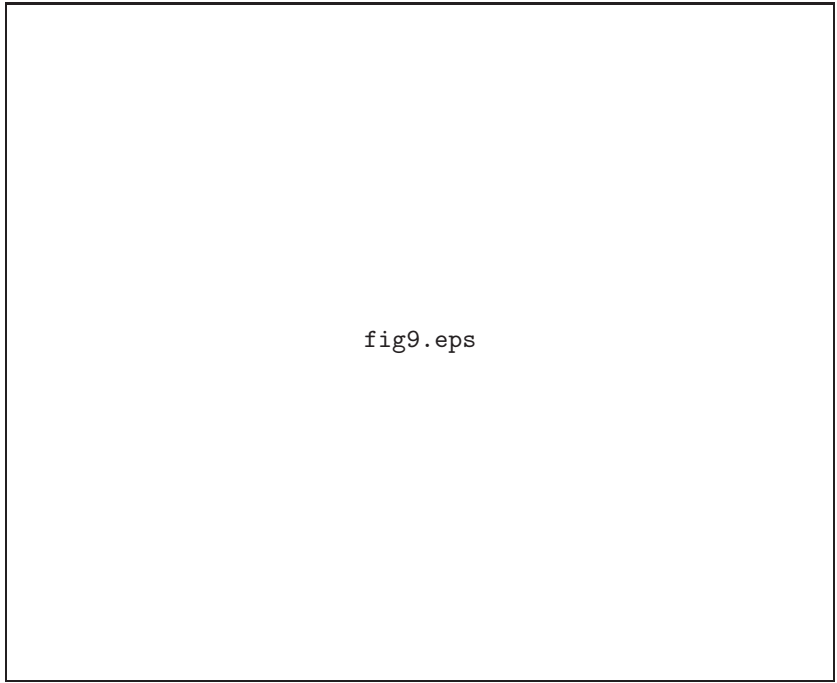

  \begin{center}
    \FigureFile(110mm,90mm){fig9.eps}
  \end{center}
  \caption{Cylindrical coordinates and relations of the rotational vectors. 
  The observer is located at the infinitely distant point on the z-axis.
 It is assumed that the system rotates around y-axis, and that
 the rotational velocity, $V$, decreases with the cosine law [proportional to 
 ($x^2+z^2)^{0.5}/R)]$. Using $V_{\rm rot,r}=V_{x} cos(\phi)$ and  $V_{\rm rot,\phi}=-V_{x} sin(\phi)$, 
 and $x=r\; cos (\phi)$ and $y=r\; sin(\phi)$, we obtain equations (24)--(26). 
  } \label{fig: Cy coordinates}
\end{figure}
\end{document}